# GPU-based Low Dose CT Reconstruction via Edge-preserving Total Variation Regularization[†]

**Zhen Tian[1,3] Xun Jia[2,3], Kehong Yuan[1], Tinsu Pan[4], Steve B. Jiang[2,3]**

[1]Department of Biomedical Engineering, Graduate School at Shenzhen, Tsinghua University, Shenzhen, Guangdong 518055, China

[2]Department of Radiation Oncology, University of California San Diego, La Jolla, CA 92093, USA

[3]Center for Advanced Radiotherapy Technologies, University of California San Diego, La Jolla, CA 92093, USA

[4]Department of Imaging Physics, University of Texas M. D. Anderson Cancer Center, Houston, Texas 77030, USA

Email: sbjiang@ucsd.edu

High radiation dose in CT scans increases a lifetime risk of cancer and has become a major clinical concern. Recently, iterative reconstruction algorithms with Total Variation (TV) regularization have been developed to reconstruct CT images from highly undersampled data acquired at low mAs levels in order to reduce the imaging dose. Nonetheless, the low contrast structures tend to be smoothed out by the TV regularization, posing a great challenge for the TV method. To solve this problem, in this work we develop an iterative CT reconstruction algorithm with edge-preserving TV regularization to reconstruct CT images from highly undersampled data obtained at low mAs levels. The CT image is reconstructed by minimizing an energy consisting of an edge-preserving TV norm and a data fidelity term posed by the x-ray projections. The edge-preserving TV term is proposed to preferentially perform smoothing only on non-edge part of the image in order to better preserve the edges, which is realized by introducing a penalty weight to the original total variation norm.

---

During the reconstruction process, the pixels at edges would be gradually identified and given small penalty weight. Our iterative algorithm is implemented on GPU to improve its speed. We test our reconstruction algorithm on a digital NCAT phantom, a physical chest phantom, and a Catphan phantom. Reconstruction results from a conventional FBP algorithm and a TV regularization method without edge preserving penalty are also presented for comparison purpose. The experimental results illustrate that both TV-based algorithm and our edge-preserving TV algorithm outperform the conventional FBP algorithm in suppressing the streaking artifacts and image noise under the low dose context. Our edge-preserving algorithm is superior to the TV-based algorithm in that it can preserve more information of low contrast structures and therefore maintain acceptable spatial resolution.

## 1. Introduction

X-ray computed tomography (CT) has been extensively used in clinics to provide patient volumetric images for a number of purposes nowadays. However, by its nature, CT scans expose a high x-ray radiation dose to the patient which may result in a non-negligible lifetime risk of cancer (Hall and Brenner, 2008; de Gonzalez *et al.*, 2009; Smith-Bindman *et al.*, 2009). This fact has become a major concern for the clinical applications of CT scans, particularly for pediatric patients, who are more sensitive to radiation and have a longer life expectancy than adults (Brenner *et al.*, 2001; Brody *et al.*, 2007; Chodick *et al.*, 2007). Therefore, it is highly desirable to reduce CT imaging dose while maintaining clinically acceptable image quality.

A simple way to reduce the x-ray dose is to lower mAs levels in CT data acquisition protocols. Nonetheless, this approach will result in an insufficient number of x-ray photons detected at imager and hence elevate the quantum noise level on the sinogram. As a consequence, the quality of the CT images reconstructed from a conventional filtered backprojection (FBP) algorithm (Deans, 1983) will be degraded by the noise-contaminated sinogram data. Another way to reduce imaging dose is to decrease the number of x-ray projections acquired by, *e.g.*, operating the x-ray generator in a high-frequency pulsed model in future with hardware modification. Yet, this will cause serious streaking artifacts in the reconstructed CT images, as the FBP algorithms require that the number of projections should satisfy the Shannon sampling theorem (Jerri, 1977).

Recently, compressed sensing algorithms (Donoho, 2006) have been applied to the CT reconstruction problem. In particular, Total Variation (TV) methods (Rudin *et al.*, 1992) have presented their tremendous power in CT reconstruction with only a few x-ray projections (Sidky *et al.*, 2006; Song *et al.*, 2007; Chen *et al.*, 2008; Sidky and Pan, 2008). In such approaches, an energy function of a TV form is minimized subject to a data fidelity condition posed by the x-ray projections. Since this energy term corresponds to image gradient, minimizing it will effectively remove those high spatial gradient parts such as noise and streaking artifacts in the reconstructed CT images. One disadvantage of this TV approach is its tendency to uniformly penalize the image gradient irrespective of the underlying image structures. As a result, edges, especially those of low contrast regions, are sometimes over smoothed, leading to the smoothed edge to some extent and loss of low contrast information. To resolve this issue, an edge guided compressive sensing reconstruction algorithm has recently been proposed in solving an MRI image reconstruction problem (Guo and Yin, 2010). In such an approach, edges are detected during the reconstruction process and are purposely excluded from being smoothed by the TV norm, by which the reconstruction can be speed up. Though this method can be easily generalized to the CT reconstruction problems, when it comes to the low dose case where insufficient or/and noisy x-ray projections are used in the reconstruction, a high quality edge detection algorithm is required to avoid incorrect detection of those fake edges caused by image noise and streaking artifacts. The sophisticated edge detection algorithms usually pose a high computation burden in the reconstructions. Moreover, they may require fine tuning of some case dependent parameters, making it hard to control their efficacy.

Another disadvantage of the TV-based reconstruction methods are the associated time consuming computation process due to their iterative nature. Generally speaking, the prolonged computational time makes the iterative CT reconstruction approaches prohibitive in many routine clinical applications. Recently, high-performance graphics processing units (GPUs) have been reported to speed up heavy duty computational tasks in medical physics, such as CBCT reconstruction (Xu and Mueller, 2005, 2007; Li *et al.*, 2007; Yan *et al.*, 2008; Jia *et al.*, 2010b), deformable image registration (Sharp *et al.*, 2007; Samant *et al.*, 2008; Gu *et al.*, 2010), dose calculation (Gu *et al.*, 2009; Hissoiny *et al.*, 2009; Jia *et al.*, 2010a) and treatment plan optimization (Men *et al.*, 2009; Men *et al.*, 2010). In principle, high computation efficiency can be expected utilizing GPU in our CT reconstruction problem.

In this work, we generalize the TV-based CT reconstruction algorithm to an edge-preserving TV (EPTV) regularization form to reconstruct CT images under undersampling and/or low mAs situation. In particular, an EPTV norm is designed by introducing a penalty weight to the original TV norm, which enables the algorithm to automatically locate the sharp discontinuities of image intensity and adjust the weight adaptively to adopt the progressively recovered edge information during the reconstruction process. This regularization term automatically ensures that less smoothing is performed on edges to better preserve the edges, as will be seen in below. Our reconstruction algorithm is implemented on GPU to speed up the computation process.

## 2. Methods

### 2.1 Reconstruction Model

CT projection can be mathematically formulated as a linear equation,

$$Pf = y, \tag{1}$$

where $f$ is a vector whose entries correspond to the x-ray linear attenuation coefficients at different voxels of the patient image. $P$ is a projection matrix in fan-beam geometry. Numerically, we generate its element $p_{i,j}$ by the length of the intersection of the x-ray $i$ with the pixel $j$, which can be explicitly computed by Siddon's fast ray tracing algorithm(Siddon, 1985). The vector $y$ represents the log-transformed projection data measured on image detectors at various projection angles. A CT reconstruction problem is formulated as to retrieve the unknown vector $f$ based on the projection matrix $P$ and the observation vector $y$. When it comes to an undersampled problem where inadequate projection data is used to reconstruct the CT image, the problem become underdetermined and there exist infinitely many solutions to the Eq. (1).

As opposed to solve the linear equation directly, the CT image can be reconstructed by minimizing an energy function with a TV regularization term:

$$f = \text{argmin}_f E(f) = \text{argmin}_f \frac{\mu}{2} \|Pf - y\|_2^2 + J(f), \ \ s.t. \ f_i \geq 0 \text{ for } \forall i, \tag{2}$$

where $\|\cdot\|_n$ denotes the $l_n$ vector norm in the imager vector space. In Eq. (2), the first term is known as a data fidelity term, which ensures the consistency between the reconstructed image $f$ and the measurement $y$. The second one is a regularization term, which is chosen to be a TV semi-norm. The introduction of the TV term in this optimization process differentiates those infinite many solutions to the Eq. (1) and picks out the one with desired image properties as the reconstructed image. Specifically, the TV term is defined as

$$J(f) = \|f\|_{TV} = \int \|\nabla f(x)\|_1 \mathrm{d}x \ , \tag{3}$$

where $\nabla f(x)$ represents the gradient of an image $f$ at a pixel $x$. The TV term has been shown to be robust to remove noise and artifacts in the reconstructed image $f$ (Sidky and Pan, 2008; Jia *et al.*, 2010b). A scalar $\mu$ is introduced to adjust the relative weights between the data fidelity term and the regularization term. This parameter depends on different factors, such as the complexity of the image content and the extent of the consistency between the measured data and the ground truth image. In this paper we choose it manually to yield a good reconstruction quality.

Despite the great success of the TV model in terms of reconstructing high quality CT images, edges around low contrast regions are sometimes oversmoothed and low contrast information is lost as a consequence. To overcome this limitation, we propose an EPTV regularization term by introducing a penalty weight vector ω in defining the TV term, namely

$$J(f) = \|f\|_{EPTV} = \int \omega(f(x)) \|\nabla f(x)\|_1 \mathrm{d}x \ . \tag{4}$$

$$\omega(f(x)) = \exp[-(\|\nabla f(x)\|_1 / \sigma)^2]. \tag{5}$$

The parameter $\sigma$ controls the amount of smoothing that we would like to apply to those pixels at edges, especially the low contrast edges, relative to those non-edge pixels. Apparently, the choice of $\sigma$ is of central importance for the algorithm. A large $\sigma$ is not able to differentiate image gradients at different pixels. In such a circumstance, the algorithm becomes essentially the TV method. In contrast, small $\sigma$ tends to give small weights to almost every pixel, making the EPTV norm inefficient in removing noise or streaking artifacts. Since the gradient values changes during the iterative reconstruction process, we propose to adaptively set the value of $\sigma$ according to the histogram of the gradient magnitude, so that a certain percentage pixels have the gradient values larger than $\sigma$. The specific percentage is dependent on the image complexity, that is, the complex images with more structures and edges should have a larger percentage than the simple images with fewer structures. In particular, for the test cases in this paper, the percentage is chosen 90%~95% as we found this choice is a good balance between reconstruction efficiency and image quality. Besides, during the reconstruction process, the pixels at edges would be gradually picked out and given small penalty weight that could be neglected, which makes the EPTV term sparser and speed up the implementation.

## 2.2 Optimization Approach

Since the cost function defined in this minimization problem is convex, it is sufficient to consider the optimality condition:

$$\mu P^T(Pf - y) + \frac{\delta}{\delta f} \|f\|_{EPTV} = 0, \tag{6}$$

where $P^T$ denotes the transposition matrix of $P$. By introducing a scalar parameter $\lambda > 0$ and a vector $v$, we can split the Eq. (6) as

$$\mu P^T(Pf - y) = \lambda(f - v),$$
$$\frac{\delta}{\delta f}\|f\|_{EPTV} = \lambda(v - f). \tag{7}$$

This inspires us that the reconstruction problem can be solved by iteratively performing the following two steps:

$$\text{(P1): } v = f - \frac{\mu}{\lambda}P^T(Pf - y),$$
$$\text{(P2): } f = \arg\min\left(\frac{\lambda}{2}\|f - v\|_2^2 + \|f\|_{EPTV}\right), \tag{8}$$

The meaning of these two steps is straightforward. In each iteration, it first obtains a trial solution in (P1) constrained by the data fidelity term. This trial solution may contain noise and artifact, since no regularization is performed. Then the subproblem (P2) utilizes the EPTV term to effectively remove those noise and artifacts, improving image quality. This two step iteration is actually the so-called  forward-backward splitting algorithm, whose mathematical properties, such as convergence, has been discussed in (Combettes and Wajs, 2005; Hale *et al.*, 2008). Xu and Mueller proposed a similar two-step reconstruction algorithm, where OS-SIRT is used to obtain a trial solution and the image enhancement is achieved by using bilateral filter or TV minimization (Xu and Mueller, 2009b).

The subproblem (P1) in Eq. (8) is a gradient descent update with a step size of $\mu/(2\lambda)$ for a minimization problem $v = \arg\min_f \|Pf - y\|_2^2$. The introduction of the parameter $\lambda$ in the algorithm is to ensure a small step size for numerical stability consideration. In this work, we modify the subproblem (P1) into using a conjugate gradient least square (CGLS) method (Hestenes and Stiefel, 1952) to solve the minimization problem instead of the gradient descent update. The skeleton of the CGLS method is presented in Appendix I. We found that the efficiency of the algorithm has been improved with this modification, yet the mathematical proof of its convergence is still needed. The minimization problem (P2) is solved with a simple gradient descent (GD) method in an iterative manner. At each iteration, the gradient direction $g$ is first numerically calculated. An inexact line search is then performed along the negative gradient direction and a step size is determined according to Amijo's rule (Bazaraa *et al.*, 2006). The solution is then updated accordingly. The penalty weight is also updated accordingly at each iteration. Specifically, the gradient direction $g$ is calculated as:

$$g = \lambda\left(f - v^{(k)}\right) + \frac{\delta}{\delta f}\|f\|_{EPTV}, \tag{9}$$

$$\frac{\delta}{\delta f}\|f\|_{EPTV} = -\nabla\frac{\omega(f(x))\nabla f(x)}{\|f(x)\|_{TV}} \tag{10}$$

Numerically, we approximate the functional variation of EPTV, $\frac{\delta}{\delta f}\|f\|_{EPTV}$, by a symmetric finite difference scheme to ensure the stability of the algorithm. The detailed derivation and the approximation of the variation of EPTV are described in Appendix II. Moreover, since physically those pixel values are x-ray attenuation coefficients and has to be positive, we have also ensure this condition by a simple truncation of those negative pixel values of the reconstructed images in each iteration. In summary, the EPTV algorithm is implemented as follows.

**EPTV Algorithm:**

Initialize $f^{(0)} = 0$. For $k = 1, 2, ...,$ do the following steps until convergence.

1. Solve (P1): $v^{(k+1)} = \text{argmin}_f \|Pf - y\|_2^2$ with an initial value $f^{(k)}$ using CGLS solver.
2. Solve (P2): $f^{(k+1)} = \text{argmin}_f \frac{\lambda}{2} \|f - v^{(k+1)}\|_2^2 + \|f\|_{EPTV}$ with GD method, and $\omega$ is updated at each iteration in this step.
3. Ensuring image positivity: $f^{(k+1)} = 0$ if $f^{(k+1)} < 0$.

The number of iterations used in our algorithm is case dependent, as it varies with the number of projections used for reconstruction, the noise level due to the mAs level used in scanning and so on. Generally, the fewer projections are used, the more obvious streak artifacts are yielded due to undersampling, the more iterations are required to suppress these streaks. Similarly, the lower mAs level is used, the more iterations are needed to remove the noise.

### 2.3 Details in CUDA implementation

Our GPU-based reconstruction code was developed under the Compute Unified Device Architecture (CUDA) programming environment and GPU hardware platform. This platform enables a number of tasks implemented in parallel on different CUDA threads simultaneously, which speeds up the performance of our reconstruction algorithm.

There are two strategies to implement the Step1 in Algorithm1 on GPU according to the size of the projection matrix $P$. For the cases with a small image size and a small amount of projections where the projection matrix $P$ and its transposition matrix $P^T$ are small enough to be stored in GPU memory, the strategy is to pre-calculate the projection matrix and save it and its transposition in GPU as a lookup table. Then CGLS method used in Step 1 becomes a set of matrix and vector operations, such as sparse matrix-vector multiplication, vector addition, scalar-vector operation and so on, which can be efficiently performed on GPU by adopting the fast GPU sparse matrix-vector multiplication (Bell and Garland, 2008) and CUBLAS Library (NVIDIA, 2008). For the cases where the projection matrix and its transposition are too large to be loaded on GPU, the strategy is to directly compute the forms $y = Pf$ and $f = P^T y$ repeatedly whenever needed. The former $y = Pf$ can be easily performed in parallel on GPU by making each thread responsible for one ray line using an improved Siddon's ray tracing algorithm(Siddon, 1985). The latter $f = P^T y$ is actually a backward projection in that it maps the projection $y$ on the detector back to the slice image $f$ by updating its pixel values along all the ray lines. It can be calculated by still using the Siddon's algorithm on GPU with each thread responsible for updating voxels along a ray line. However, this operation would cause a memory writing conflict problem due to the possibility of simultaneously updating a same pixel value by different GPU threads. When this conflict occurs, one thread will have to wait until other threads finish updating. It is this fact that severely limits the maximal utilization of GPU's massive parallel computing power. To solve this issue, a GPU-friendly backward-projection method is developed as follows:

$$[P^T y](u, v) = \frac{\Delta u \Delta v}{\Delta \beta} \cdot \sum_\theta \frac{y^\theta(\beta^*)}{l(u^*, v^*)}, \tag{11}$$

where $(u^*, v^*)$ is one pixel on the reconstructed CT slice, and $l(u^*, v^*)$ is the distance from the x-ray source to the pixel $(u^*, v^*)$. $\beta^*$ is the angular coordinate of the rayline connecting the x-ray source and the pixel $(u^*, v^*)$ at the projection angle $\theta$. $y^\theta(\beta^*)$ is the corresponding detector reading. $\Delta u$ and $\Delta v$ are the size of pixels on CT slices, and $\Delta \beta$ is the angular spacing of the detector units. The derivation of Eq. (11) is briefly shown in the Appendix III. By this backward projection method, we can simply get the value of $f$ at a given pixel $(u^*, v^*)$ by summing the projection values $y^\theta(\beta^*)$ over all projection angles $\theta$ after proper geometry correction, which allows us to implement the calculation in parallel on GPU with each thread responsible for one pixel and thus to avoid the memory writing conflict problem. In numerical computation, since we always evaluate $y^\theta(\beta)$ at a set of discrete coordinates and $\beta^*$ does not necessarily coincide with these discrete coordinates, a linear interpolation is performed to obtain $y^\theta(\beta^*)$.

The main operations in the Step 2 are calculating the gradient direction, descending along the direction, and updating the penalty weight *etc.* All of them can be easily implemented in parallel on GPU with each thread computing one entry. The Step 3 can be similarly processed. The flow chart of our reconstruction algorithm is shown in Fig. 1.

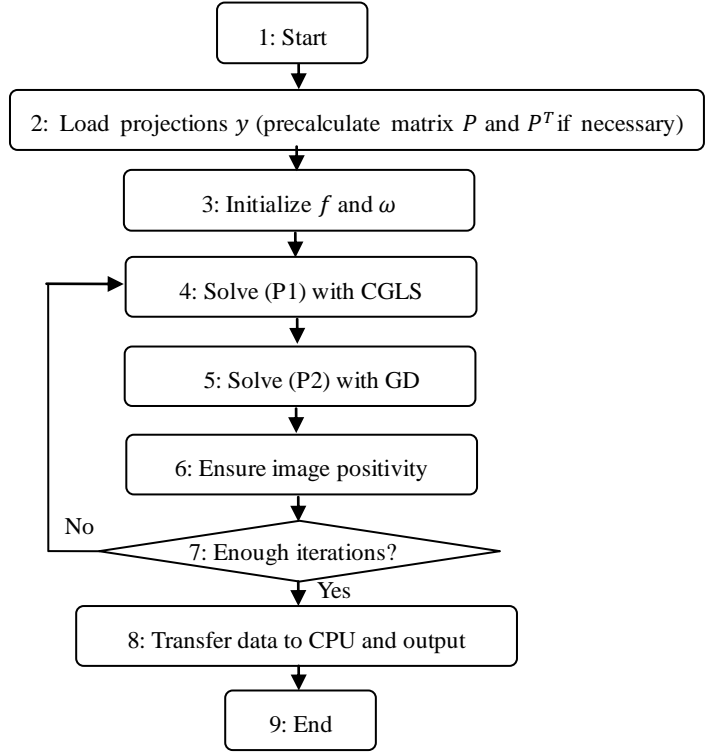

**Figure 1**. The flow chart of our GPU-based EPTV reconstruction. Blocks 4~6 correspond to the Steps 1~3 of the EPTV Algorithm.

## 3. Experimental Results

We tested our reconstruction algorithm on three cases: a digital NURBS-based cardiac-troso (NCAT) phantom in thorax region (Segars, 2002), a physical chest phantom, and a Catphan phantom. In digital phantom experiments, we simulated x-ray projections using Siddon's ray tracing algorithm (Siddon, 1985) in fan-beam geometry with an arc detector of 888 units and a spacing of 1.0239mm. The source to detector distance is 949.075mm and the source to rotation center distance is 541.0mm. All of these parameters mimic a realistic configuration of a GE Lightspeed QX/I CT scanner. In all cases, we simulated/acquired x-ray projection data along 984 directions equally spaced in a full rotation and a subset of them were used for reconstruction. The size of reconstructed images is 512×512.

*3.1 Digital phantom experiment*

We first tested our EPTV algorithm on an NCAT phantom in thorax region using undersampled x-ray projections. In order to better illustrate how our EPTV algorithm performs, in Fig. 2 we show the penalty weights and the reconstructed image at different iterations during the reconstruction with 40 projections being used. During the evolution, the algorithm automatically picked out the edges and gave them small weight *w* in the EPTV term to avoid being oversmoothed, shown in Fig. 2 (2nd row). Accordingly, the structural information is gradually resolved in the reconstructed image, with main structures and obvious artifacts at first and low contrast structures and fewer artifacts later, shown in Fig.2 (4th row). Note that there are some false edges detected at iteration 5, which are gradually reduced with more iterations, leading to visually better reconstructed images.

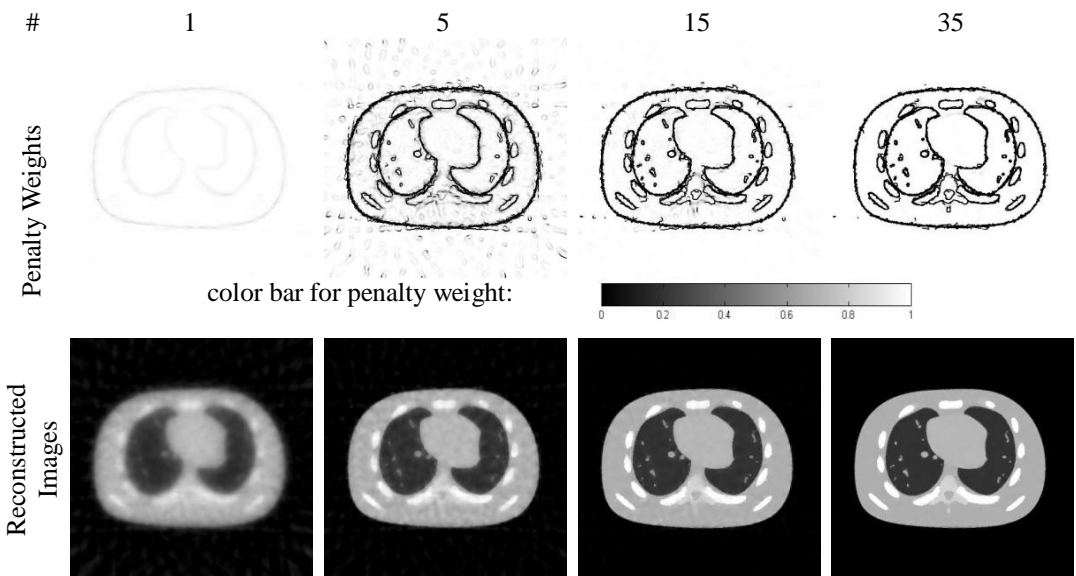

**Figure 2.** The evolution of the penalty weights and reconstructed image at different numbers of iteration (top row) during the reconstruction using our EPTV algorithm with 40 projections.

The reconstructed images using 40 projections with 50 iterations are shown in Fig. 3. We can see that the conventional FBP algorithm is not able to reconstruct the CT images with such few projections and obvious streaking artifacts are observed, which makes the image unacceptable. In contrast, even with such few projections, both the TV method and the EPTV method can still capture most of the structures, leading to visually much better reconstruction results. As for the comparison between the TV method and the EPTV method, the spinal bone structures of low contrast indicated by the arrow in Fig. 3(c) and (d) are hardly resolved by the TV method, while much clearer structures with sharper edges are observed in the image reconstructed by the EPTV method. To better compare the TV method and the EPTV method in detail, we also show the horizontal and vertical intensity profiles going through the center of the reconstructed images. Clearly, large fluctuations at non-edge points are observed in the profiles obtained from the TV method. Besides, the jumps indicated by arrows are broadened and weakened, indicating blurring effects and losing of contrast. On the other hand, the profiles of EPTV are very close to that of the ground truth with sharp jumps at edge points and small fluctuation at non-edge points. All of these clearly demonstrate the advantages of the EPTV algorithm in resolving the low contrast edges over the TV algorithm within the same number of iterations.

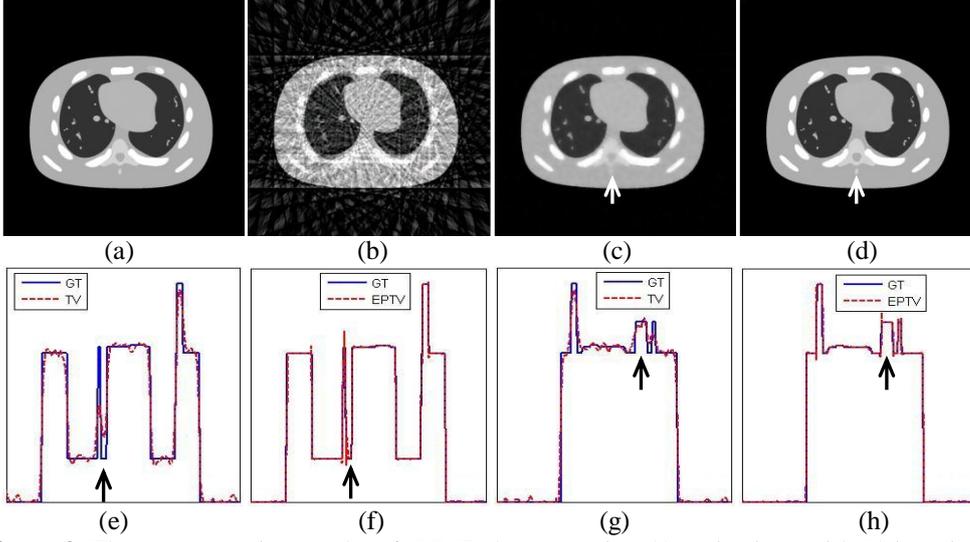

**Figure 3.** The reconstruction results of NCAT phantom using 40 projections with 50 iterations. (a) is the ground truth image. (b) ~ (d) show the images reconstructed by FBP, TV, EPTV, respectively. (e)(f) depict the horizontal intensity profiles through the center of reconstructed images of TV and EPTV, while (g)(h) are the corresponding vertical intensity profiles. The profiles of ground truth (GT) are also plotted in solid lines for comparison. The arrows indicate the area where EPTV is clearly superior to TV.

To quantify the reconstruction accuracy of the algorithm, we use relative error as a metric to measure the similarity between the ground truth image and reconstructed image. Besides, since edge information is an important feature of images and human vision is highly sensitive to it, edge cross-correlation coefficient (ECC) (Xu and Mueller, 2009a) is also used as another metric to evaluate the reconstruction accuracy. These metrics are calculated as follows:

$$e = \frac{\|f - f^*\|_2^2}{\|f^*\|_2^2} \tag{12}$$

$$c = \frac{\sum_i (b_i - \bar{b})(b_i^* - \bar{b}^*)}{\sqrt{\sum_i (b_i - \bar{b})^2 \sum_j (b_i^* - \bar{b}^*)^2}} \tag{13}$$

where $f^*$ is the ground truth image and $b$ is a binary image with 1 indexing image edges on the reconstructed image and 0 otherwise. The edges are detected using a standard Sobel edge detector (Kanopoulos *et al.*, 1988). $b^*$ is defined in a similar manner for the ground truth image. The over-bar indicates an average of the corresponding quantities over all pixels.

The relative error $e$ and the edge correlation coefficient $c$ of the reconstructed images of the digital phantom are shown as functions of number of projections and number of iterations, respectively, in Fig. 4. As expected, the more projections used, the better reconstruction quality will be obtained with smaller relative error and higher edge correlation coefficient. Fig. 4(a) and (b) clearly demonstrate that with the same number of projections, the EPTV method is superior to the TV method with a smaller error and higher edge cross-correlation. In this particular case, 40 projections are sufficient for EPTV reconstruction to clearly resolve the low-contrast structures. The evolution curves of the relative error and the edge correlation during iterations in Fig. 4(c) and (d) show that the EPTV algorithm converges much faster than the TV method. With these curves, we can conclude that the EPTV method outperforms the TV method, in that it leads to smaller relative error, higher correlation, and faster convergence for a given number of projections.

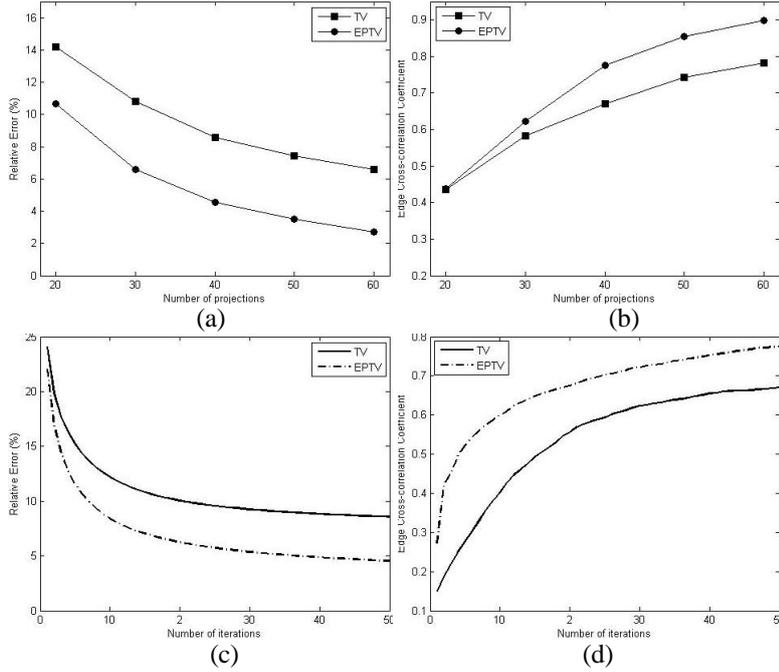

**Figure 4.** (a)(b) show the relative error and the edge cross-correlation coefficient as functions of the number of x-ray projections at the 50th iteration; (c)(d) show the evolution curve of the relative error and the edge cross correlation during the first 50 iterations using 40 projections.

### 3.2 Physical phantom experiments

In this paper, we scanned both a physical chest phantom and a physical Catphan phantom under high dose and low dose protocols. The experiment on the chest phantom was designed to evaluate the tolerance of the EPTV algorithm to image noise due to low imaging dose, while the experiment on the Catphan phantom aimed to evaluate the performance of our algorithm in terms of spatial resolution.

In the experiment on the chest phantom, for the high dose protocol, 984 projections with 0.4mAs/projection were acquired and then used to reconstruct the image with the FBP algorithm. In the low dose protocol, the chest phantom was scanned at 0.02mAs/projection and only 200 equally spaced projections are used for reconstruction. The dose in the low dose protocol is about 1% of that in the high dose protocol. The reconstruction results with 30 iterations are shown in Fig. 5. Two different display windows are used to stress the bones and the details in the lung, respectively. We can see that using FBP in such a low dose situation, the image quality deteriorates greatly with obvious streaking artifacts and image noise. In contrast, images reconstructed by TV and EPTV under the same low dose context show better image quality with less noise and streaking artifacts. While comparing TV and EPTV methods, the latter is found to be superior on preserving the edges of fine contrast structures. Take the area indicated by the arrows as an example, the bone is much clearer with a hollow in the image reconstructed by EPTV than that in the image reconstructed by TV. Although the cloud-like and low contrast details in the lung region in Fig. 5(b1) are great challenges to be reconstructed for both the TV algorithm and the EPTV algorithm, they are better preserved by the latter.

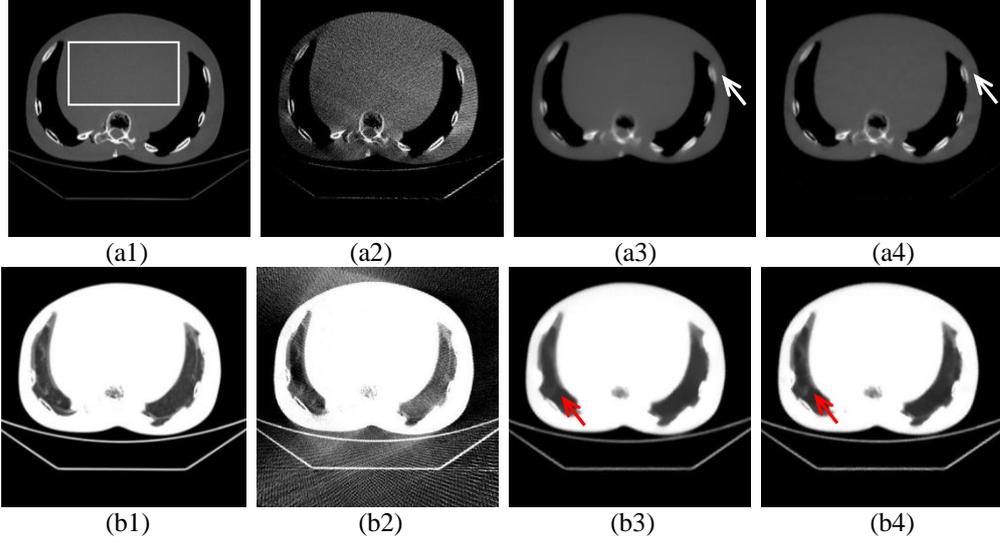

<div align="center">(a1)        (a2)        (a3)        (a4)</div>

<div align="center">(b1)        (b2)        (b3)        (b4)</div>

**Figure 5.** Reconstruction results of a physical chest phantom with 30 iterations. (1) shows the image reconstructed by FBP at 400mAs (984 projections × 0.407mAs/projection). (2), (3), and (4) show the images reconstructed by FBP, TV, EPTV, respectively, at 4mAs (200 projections × 0.0203mAs/projection). The CT images are shown with display window [-400 1000] and [-900 10] in the two rows to show the bones and the details in the dark lung region, respectively. The rectangle shows ROI used for SNR calculation. Arrows indicate the area where EPTV is clearly superior to TV.

To quantify the image quality of reconstructed images in terms of image noise, the signal to noise ratio (SNR) of a region of interest (ROI) is shown in Table 1. The calculation is as follows:

$$SNR = 10 \times \log_{10}\left[\frac{E^2(P)}{E([P-E(P)]^2)}\right], \tag{14}$$

where $p$ is the set of pixels inside the ROI and $E(\cdot)$ stands for a spatial average over the ROI. It can be observed from Table 1 that when the imaging dose is decreased by 100 times, the SNR of images reconstructed by FBP is lowered from 40.6 to 20.4, indicating the sensitivity of the FBP method to image noise and its inability to handle the low dose situation. In contrast, the SNRs of images reconstructed by TV and EPTV methods are only decreased by a little compared to images reconstructed by FBP with high dose. With these results, we can conclude that our EPTV method not only outperforms the TV method on preserving the low contrast edges, but also inherits the advantage of the TV method on handling the low dose situation and effectively suppressing image noise.

|     | FBP  | FBP  | TV   | EPTV |
|-----|------|------|------|------|
| mAs | 400  | 4    | 4    | 4    |
| SNR | 40.6 | 20.4 | 39.9 | 40.0 |

**Table 1**. SNR of the ROI in the reconstructed images of the chest phantom.

The experimental results of the Catphan phantom with 20 iterations are shown in Fig. 6.

The low imaging dose in this case is 8mAs, about 2% of that in the high dose scan, by decreasing the number of projections from 984 to 400 and reducing the mAs/projection from 0.4 to 0.02 mAs/projection. The sixth line pair at which the arrows point clearly demonstrate the different performances of the three reconstruction algorithms in terms of resolution. For the FBP reconstruction at low dose, the sixth line pair can still be clearly resolved as in Fig. 6(f). Compared with the reconstructed image of high dose, the low dose image reconstructed by FBP deteriorates mainly due to image noise, but still maintains good spatial resolution (6.5 lp/cm). On the other hand, although the TV method performs well in suppressing the image noise, the spatial resolution is degraded to 4 lp/cm. The sixth pair of lines is blurred and cannot be visually resolved, see Fig. 6(g). Finally, for the image reconstructed by the EPTV method, it contains less noise and maintains high spatial resolution (6.5 lp/cm), where the lines in the sixth pair can be clearly identified from each other in Fig. 6(h). The results of this test confirm the conclusion that, not only does our EPTV method effectively reduce image noise in low dose CT reconstruction, but also performs better than TV in maintaining acceptable spatial resolution. However, the EPTV algorithm prefer to perform less smoothing on edges, which makes the edges not as smooth as those reconstructed by the TV algorithm. Therefore, the EPTV algorithm maintains higher spatial resolution at the cost of the smoothness of the edges.

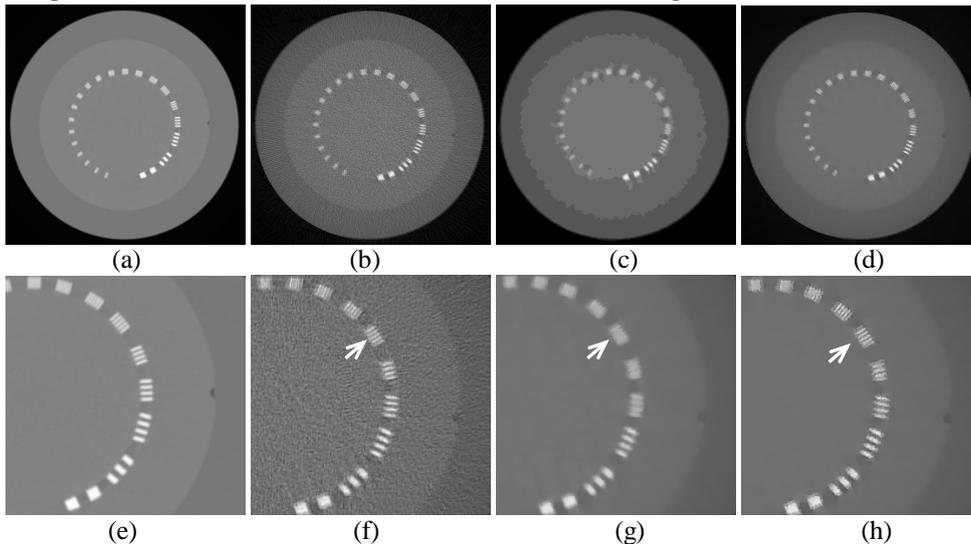

**Figure 6.** Reconstruction results of the Catphan phantom with 20 iterations. (a) shows the image reconstructed by FBP at 400mAs (984 projections × 0.407mAs/projection). (b), (c), and (d) show the images reconstructed by FBP, TV, EPTV, respectively, at 8mAs (400 projections × 0.0203mAs/projection). (e)-(h) zoom in the corresponding images on the first row. Arrows indicate the area where EPTV is clearly superior to TV.

*3.3    Computation efficiency*

To speed up our iterative algorithm, we have implemented and tested it on an NVIDIA Tesla C1060 card. For comparison purpose, the algorithm in C language was also implemented on a 2.27 GHz Intel Xeon processor. The computation times of the EPTV algorithm on CPU and GPU for an axial CT slice are shown in Table. 2. The absolute computation time is case dependent as it varies with the number of projections used ($NP$), the number of iterations ($NI$), *etc.* However, when comparing $T_{CPU}$ and $T_{GPU}$ for a same case, it is found that the algorithm is sped up by factors of about 19.6~23.1. This considerable enhancement of efficiency implies a potential application of the EPTV algorithm in a realistic clinical environment to reduce the imaging dose. Although currently the reconstruction time still cannot compete with that of the conventional FBP, it is hoped that multi-GPU would be used in future to implement the reconstruction for multiple slices simultaneously, which would speed up the algorithm further. We will also optimize the algorithm to further improve its efficiency.

|         | NP  | NI | $T_{CPU}$ (s) | $T_{GPU}$ (s) | $T_{CPU}/T_{GPU}$ |
|---------|-----|----|---------------|---------------|-------------------|
| NCAT    | 40  | 50 | 126.9         | 5.5           | 23.2              |
| Chest   | 200 | 30 | 212.3         | 9.8           | 21.7              |
| Catphan | 400 | 20 | 374.8         | 19.1          | 19.6              |

**Table 2.** The computation time of the EPTV algorithm on CPU and GPU.

## 4. Discussion and Conclusions

Two possible ways to reduce the imaging dose of CT scans are to reduce the number of projections and lower the mAs per projection. However, the common FBP reconstruction algorithms result in streaking artifacts and obvious image noises in these circumstances, which can be clearly observed in the experimental results. The iterative algorithm with TV norm as a regularization term has presented its tremendous power in CT reconstruction with undersampled projections acquired at low mAs level. However, the low contrast structures tend to be smoothed out by the TV regularization, posing a great challenge to the TV method. In this work, we propose an edge-preserving TV norm to solve the problem by giving adaptive penalty weights to pixels to suppress smoothing near both the strong edges and low contrast edges, relative to the non-edge pixels. By this adaptive penalization mechanism, our algorithm could automatically pick out the edges gradually during the implementation and give them a small weight and thus preferentially performs smoothing on those non-edge pixels to reduce image noise while keeping those edges unaffected, especially the low contrast edges. Our EPTV approach has been validated on a digital NCAT phantom, a physical chest phantom and also a Catphan phantom. The experimental results have clearly demonstrated the advantage of our approach over the conventional filtered backprojection algorithm and the TV-based reconstruction algorithm under low dose context on effectively suppressing image noise and streaking artifact and also maintaining acceptable signal-to-noise ratio and spatial resolution, which indicates a promising prospect of our low dose CT reconstruction algorithm in clinical applications for a considerable reduction of radiation dose exposed to patients.

To further understand the advantages of our EPTV algorithm over the TV algorithm, it is better to consider how the minimization is performed for the problem (P2) in Eq. (8). This model essentially tend to enhance image quality of an input image $v$ by utilizing a regularization function $J[f]$, which is of a total variation form in the conventional TV method and EPTV term in this paper. It is straightforward to perform the minimization by using a gradient descent approach, or mathematically equivalently to solve a partial differential equation (PDE), $\frac{\partial f}{\partial t} = -\lambda(f - v) - \frac{\partial J[f]}{\partial f}$, where $t$ is a variable to parameterize the evolution of the solution. In TV case, the variation term on the right hand side is $\frac{\partial J[f]}{\partial f} = -\nabla \cdot (\frac{1}{|\nabla f|} \nabla f)$. This equation is then essentially a diffusion equation with a diffusion

constant $\frac{1}{|\nabla f|}$. During the evolution according to the PDE, the high frequency signals, such

as noise, can be removed due to the diffusive process. Meanwhile, sharp edges are
preserved, as at those sharp edges $|\nabla f| \to \infty$ and hence the diffusion constant approaches

5    zero. It is essentially for this reason that TV method can be very effective in terms of
removing noise while preserving sharp edges. In a continuous model, apparently sharp
edges will be perfectly preserved since the diffusion constant becomes exactly zero at edges.
However, in a discrete case as in numerical calculation, the diffusion constant will never be
exactly zero due to the use of the finite difference. This means the diffusion process during

10   the optimization will more or less take place at edges no matter how sharp the edge is. This
will lead to blurring of the edge to a certain extent, though very small for sharp edges. This
is even worse for low contrast edges, as the diffusion is significant. As for our EPTV

algorithm, we can perform the same analysis and obtain $\frac{\partial J[f]}{\partial f} = -\nabla \cdot (\frac{w}{|\nabla f|} \nabla f)$. The

diffusion constant now becomes $\frac{w}{|\nabla f|}$, where it is always true that $w < 1$ according to its

15   definition. Therefore, the diffusion at edges is always smaller in EPTV than in TV, leading
to better edge preserving. Moreover, during the optimization, the parameter $w$ is adaptively
adjusted so that the distinction between edge and non-edge are always kept, making the
algorithm to preferentially perform smoothing on those non-edge pixels. Besides, note that
TV term is a special case of the EPTV term with all $w$ factors equal to unity and it holds

20   that $w < 1$ in the EPTV case, therefore the EPTV term is smaller than the corresponding
TV term, which makes the EPTV term a sparser representation than the original TV term.

GPU is employed in our experiments to improve the speed of this iterative algorithm,
which is always a big challenge for the clinical applications of this type of algorithms. It is
found that the GPU implementation is able to speed up the iterative reconstruction process

25   by factors of 19.6~23.2 and make it possible to reconstruct an image in seconds instead of
minutes. Though currently the reconstruction speed still cannot compete with the speed of
the conventional FBP, it is hoped that the efficiency can be improved by using multi-GPU
and reconstructing multiple CT slices simultaneously in the near future. We would also
work on optimizing our algorithm to improve the speed further.

30   The choice of the parameter $\mu$ is of central importance to the success of this EPTV
reconstruction method. In this preliminary study, we manually choose this parameter to
adjust the relative weight between the data fidelity term and the regularization term to get
the best reconstructed CT image quality. It is found that the optimal value of this parameter
is case dependent. In future, we would study on the parameter setting try to find an

35   automatic or semi-automatic way to guarantee a good choice of the parameter.

Lowering mAs level to reduce the patient radiation dose can easily be to done on
existing commercial CT scanners. While, the other method proposed in our paper, namely,
reducing the number of projections to reduce the dose is not straightforward on currently
available commercial CT scanners due to the use of continuous x-ray generation mode.

40   However, technically it is possible to modify the scanners to operate in high-frequency pulsed mode(Kang *et al.*, 2010; Myagkov *et al.*, 2009; Yue *et al.*, 2002), if there is a clinical need (such as the one suggested in this paper).

**Acknowledgements**

5    This work is supported in part by the University of California Lab Fees Research Program. The authors would like to thank NVIDIA for providing GPU cards for this project.

**Appendix**

Appendix I. CGLS algorithm

The conjugate gradient least square (CGLS) method used to solve the minimization problem
$v = \mathrm{argmin}_f \|Pf - y\|_2^2$ , which is a substitution of the subproblem (P1) in Eq. (8) , is
implemented as follows.

| **CGLS Algorithm :** |
|---|
| Initialinize: $m = 0$, $u^{(0)} = f^{(k)}$, $r^{(0)} = y - Pu^{(0)}$, $s^{(0)} = P^T r^{(0)}$. Do the Steps 1-5 for M times. |

1. $a^{(m)} = \|P^T r^{(m)}\|_2^2 / \|PS^{(m)}\|_2^2$
2. $u^{(m+1)} = u^{(m)} + a^{(m)} s^{(m)}$
3. $r^{(m+1)} = r^{(m)} - a^{(m)} PS^{(m)}$
4. $b^{(m)} = \|P^T r^{(m+1)}\|_2^2 / \|P^T r^{(m)}\|_2^2$
5. $s^{(m+1)} = P^T r^{(m+1)} + b^{(m)} s^{(m)}$
6. $v^{(k+1)} = u^{(M)}$

Appendix II . Derivation of the variation of EPTV and its finite difference scheme

To derive the functional variation of EPTV, $\frac{\delta}{\delta f} \|f\|_{EPTV}$, let us consider the integrations by
parts of the EPTV term:

$$
\begin{aligned}
\|f\|_{EPTV} &= \int \omega(f(x)) \|\nabla f(x)\|_1 \mathrm{d}x \\
&= \int \nabla f(x) \cdot \frac{\omega(f(x)) \nabla f(x)}{\|\nabla f(x)\|_1} \mathrm{d}x \\
&= f \frac{\omega(f(x)) \nabla f(x)}{\|\nabla f(x)\|_1} |_{\partial\Omega} - \int f \cdot \nabla \frac{\omega(f(x)) \nabla f(x)}{\|\nabla f(x)\|_1} \mathrm{d}x
\end{aligned} \tag{A1}
$$

where $\Omega$ is the image domain and $\partial\Omega$ is its boundary. Assume zero boundary condition, the
first term vanishes. Taking variation of the second term with respect to $f$ leads to Eq. (10).
Numerically, we approximate the functional variation of EPTV by a symmetric finite
difference scheme ：

$$
\begin{aligned}
\frac{\delta}{\delta f_{i,j}} \|f\|_{EPTV} \approx &- \left[ \frac{\omega_{i+1/2,j}(f_{i+1,j} - f_{i,j})}{\|f_{i+1/2,j}\|_{TV}} - \frac{\omega_{i-1/2,j}(f_{i,j} - f_{i-1,j})}{\|f_{i-1/2,j}\|_{TV}} \right] \\
&- \left[ \frac{\omega_{i,j+1/2}(f_{i,j+1} - f_{i,j})}{\|f_{i,j+1/2}\|_{TV}} - \frac{\omega_{i,j-1/2}(f_{i,j} - f_{i,j-1})}{\|f_{i,j-1/2}\|_{TV}} \right]
\end{aligned} \tag{A2}
$$

where,

$$
\begin{aligned}
\|f_{i+1/2,j}\|_{TV} &= \left[ (f_{i+1,j} - f_{i,j})^2 + \frac{1}{2}(f_{i+1/2,j+1} - f_{i+1/2,j})^2 \right. \\
&\qquad \left. + \frac{1}{2}(f_{i+1/2,j} - f_{i+1/2,j-1})^2 \right]^{1/2}, \\
f_{i+1/2,j} &= \frac{1}{2}(f_{i+1,j} + f_{i,j}), \\
\omega_{i+1/2,j} &= \exp\left(-\frac{\|f_{i+1/2,j}\|_{TV}^2}{\sigma^2}\right).
\end{aligned} \tag{A3}
$$

Such an approximation scheme, though cumbersome, ensures the stability of the algorithm.

Appendix III. Derivation of Eq. (11)

Let $f(.): \mathbf{R}^2 \to \mathbf{R}$ and $g(.): \mathbf{R} \to \mathbf{R}$ be two smooth enough functions in the CT slice domain and in the x-ray projection domain, respectively. The operator $P^{\theta^T}$, being the adjoint operator of the x-ray projection operator $P^\theta$, should satisfy the condition

$$\langle f, P^{\theta^T} y \rangle = \langle P^\theta f, y \rangle, \tag{A3}$$

5    where $\langle . , . \rangle$ denotes the inner product. This condition can be explicitly expressed as

$$\int du\, dv\, f(u, v)[P^{\theta^T} y](u, v) = \int d\beta\, [P^\theta f](\beta) y(\beta). \tag{A4}$$

Now take the functional variation with respect to $f(x)$ on both sides of equation (A4) and interchange the order of integral and variation on the right hand side. This yields

$$[P^{\theta^T} y](u, v) = \frac{\delta}{\delta f} \int d\beta\, [P^\theta f](\beta) y(\beta) = \int d\beta\, y(\beta) \frac{\delta}{\delta f} [P^\theta f](\beta). \tag{A5}$$

With help of a delta function, the forward projection can be written as

$$[P^\theta f](\beta) = \int dl\, f(u, v)\, \delta(u - u_S - l\cos(\beta + \theta))\, \delta(v - v_S - l\sin(\beta + \theta)). \tag{A6}$$

where, $(u_S, v_S)$ is the coordinate of the x-ray source, $l$ is the distance between the source

10    and the pixel $(u, v)$ in the CT slice domain. Now substituting (A6) into (A5), we obtain

$$[P^{\theta^T} y](u, v) = \int dl\, d\beta\, y(\beta)\, \delta(u - u_S - l\cos(\beta + \theta))\, \delta(v - v_S - l\sin(\beta + \theta))$$

$$= \int dl\, d\beta\, y(\beta)\, \delta(\beta - \beta^*) \delta(l - l^*) \frac{1}{\left|\frac{\partial(h_1, h_2)}{\partial(\beta, l)}\right|_{\beta^*, l^*}} = \frac{y^\theta(\beta^*)}{l^*}. \tag{A7}$$

$\beta^*$ is the angular coordinate of the rayline connecting the x-ray source and the pixel $(u^*, v^*)$ at the projection angle $\theta$. $y^\theta(\beta^*)$ is the corresponding detector reading. Changing from continuous coordinate to discrete coordinate, the ratio of pixel size $\Delta u \Delta v$ on CT slice to the angular spacing size $\Delta \beta$ of the detector unit is added to the Eq. (A7). Additionally, a

15    summation over projection angles $\theta$ is performed to account for all the x-ray projection images. So we arrive at the Eq. (11). We have tested the accuracy of such defined operator $P^T$ in terms of satisfying condition expressed in Eq. (A3). Numerical experiments indicate that this condition is satisfied with numerical error less than 1%, which is found accurate enough for our iterative CT reconstruction purpose.